\documentclass[graybox]{svmult}
\usepackage{type1cm}        

\usepackage{makeidx}         
\usepackage{graphicx}        
\usepackage{multicol}        
\usepackage[bottom]{footmisc}

\usepackage{subcaption}
 \usepackage{acronym}

\acrodef{adc}[ADC]{Analog-to-Digital Convertor}
\acrodef{dac}[DAC]{digital-to-analog convertor}
\acrodef{cs}[CS]{Compressed Sensing}
\acrodef{em}[EM]{ElectroMagnetic}
\acrodef{dtft}[DTFT]{discrete-time Fourier transform}
\acrodef{dnn}[DNN]{deep neural network} 
\acrodef{csi}[CSI]{Channel State Information}
\acrodef{map}[MAP]{maximum a-posteriori probability}
\acrodef{snr}[SNR]{Signal-to-Noise Ratio}
\acrodef{sinr}[SINR]{signal-to-interference-and-noise ratio}
\acrodef{bs}[BS]{Base Station} 
\acrodef{iot}[IOT]{Internet of Things}
\acrodef{mimo}[MIMO]{Multiple-Input Multiple-Output}
\acrodef{mse}[MSE]{Mean-Squared Error}
\acrodef{pdf}[PDF]{probability density function}
\acrodef{rv}[RV]{random variable}
\acrodef{tdd}[TDD]{time division duplexing}
\acrodef{rs}[RS]{Reed-Solomon}
\acrodef{lti}[LTI]{linear time-invariant}
\acrodef{wss}[WSS]{wide-sense stationary}
\acrodef{psd}[PSD]{power spectral density}
\acrodef{ser}[SER]{symbol error rate} 
\acrodef{ber}[BER]{bit error rate} 
\acrodef{isi}[ISI]{intersymbol interference}  
\acrodef{awgn}[AWGN]{additive white Gaussian noise} 
\acrodef{ut}[UT]{User Terminal} 
\acrodef{dc}[DC]{Direct Current} 
\acrodef{aoa}[AoA]{Angle of Arrival} 
\acrodef{mmw}[mmWave]{millimeter wave}
\acrodef{ris}[RIS]{Reconfigurable Intelligent Surface} 
\acrodef{hris}[HRIS]{Hybrid Reconfigurable Intelligent Surface} 
\acrodef{dma}[DMA]{Dynamic Metasurface Antenna}

\usepackage{algorithm, algorithmic}
\usepackage{cite}
\usepackage{epsf,graphics,graphicx}
\usepackage{comment}
\usepackage{bm}
\usepackage{amssymb}
\usepackage{amsmath,flushend}

\usepackage{circuitikz}
\usetikzlibrary{positioning}
\usepackage{tikz}
\usepackage{ellipsis}
\usetikzlibrary{calc}
\usetikzlibrary{decorations.pathreplacing,decorations.markings,shapes.geometric}
\tikzset{naming/.style={align=center,font=\small}}
\tikzset{antenna/.style={insert path={-- coordinate (ant#1) ++(0,0.25) -- +(135:0.25) + (0,0) -- +(45:0.25)}}}
\tikzset{station/.style={naming,draw,shape=dart,shape border rotate=90, minimum width=10mm, minimum height=10mm,outer sep=0pt,inner sep=3pt}}
\tikzset{mobile/.style={naming,draw,shape=rectangle,minimum width=12mm,minimum height=6mm, outer sep=0pt,inner sep=3pt}}
\tikzset{radiation/.style={{decorate,decoration={expanding waves,angle=90,segment length=4pt}}}}

\usepackage{comment}
\usepackage[textwidth=30mm]{todonotes}
\usepackage{soul}

\setstcolor{magenta}
\sethlcolor{lightmauve}

\definecolor{deepmagenta}{rgb}{0.8, 0.0, 0.8}
\definecolor{lightmauve}{rgb}{0.86, 0.82, 1.0}
\definecolor{green-yellow}{rgb}{0.68, 1.0, 0.18}
\definecolor{lightskyblue}{rgb}{0.53, 0.81, 0.98}
\definecolor{beaublue}{rgb}{0.74, 0.83, 0.9}

\def\vec#1{{\bf #1}}
\def\Tr#1{\mathrm{Tr}\left\{ #1 \right\} }

\DeclareMathOperator{\ex}{\mathbb{E}}

\def\boldmu{\mbox{\boldmath$\mu$}}

\def\bLambda{\mbox{\boldmath$\Lambda$}}
\def\bPhi{\mbox{\boldmath$\Phi$}}
\def\bSigma{\mbox{\boldmath$\Sigma$}}

\def\bG{{\bf G}}

\def\bI{{\bf I}}
\def\bQ{{\bf Q}}
\def\bR{{\bf R}}
\def\bS{{\bf S}}
\def\bT{{\bf T}}

\def\nt{{N_t}}
\def\nr{{N_r}}
\def\ns{{N_s}}

\def\nue{{M}} 
\def\nris{K} 

\def\snrdir{{\gamma_{dm}}} 

\DeclareMathAlphabet\mathbfcal{OMS}{cmsy}{b}{n}

\def\bk{{\bf k}}

\def\bq{{\bf q}}

\def\bs{{\bf s}}

\def\bu{{\bf u}}
\def\bv{{\bf v}}

\def\bx{{\bf x}}
\def\by{{\bf y}}
\def\bz{{\bf z}}

\usepackage{newtxtext}       %
\usepackage{newtxmath}       

\makeindex             

\begin{document}

\title*{Multi-RIS-Empowered Communication Systems: Capacity Analysis and Optimization}
\author{Aris L. Moustakas 
and George C. Alexandropoulos}
\authorrunning{A. L. Moustakas and G. C. Alexandropoulos} 
\institute{Aris L. Moustakas \at Department of Physics, National and Kapodistrian University of Athens,  Greece, \email{arislm@phys.uoa.gr}
\and George C. Alexandropoulos \at Department of Informatics and Telecommunications, National and Kapodistrian University of Athens,  Greece, \email{alexandg@di.uoa.gr}
}
%
%
\maketitle

\abstract{In this chapter, using statistical physics methods, asymptotic closed-form expressions for the mean and variance of the mutual information for a multi-antenna transmitter-receiver pair in the presence of multiple Reconfigurable Intelligent Surfaces (RISs) are presented. While nominally valid in the large-system limit, it is shown that the derived Gaussian approximation for the mutual information can be quite accurate, even for modest-sized antenna arrays and metasurfaces. The above results are particularly useful when fast-fading conditions are present, which renders channel estimation challenging. The derived analysis indicates that, when the channel close to an RIS is correlated, for instance due to small angle spread which is reasonable for wireless systems with increasing carrier frequencies, the communication link benefits significantly from statistical RIS optimization, resulting in gains that are surprisingly higher than the nearly uncorrelated case. More importantly, the presented novel asymptotic properties of the correlation matrices of the impinging and outgoing signals at the RISs can be deployed to optimize the metasurfaces without brute-force numerical optimization. The numerical investigation demonstrates that, when the desired reflection from any of the RISs departs significantly from geometrical optics, the metasurfaces can be optimized to provide robust communication links, without significant need for their optimal placement.}


\section{Introduction}
The upcoming sixth Generation (6G) of wireless networks is envisioned to connect the physical, digital, and human worlds, enabling ubiquitous wireless intelligence and extremely demanding communications, sensing, and computing applications (e.g., holographic imaging~\cite{Gong_HMIMO_2023}, digital twinning~\cite{Masaracchia_DT_2023}, and edge artificial intelligence~\cite{Letaief_edgeAI_2023,RIS_OTA_EI}). This vision necessitates advances at various aspects of the network design, including the orchestration of its diverse components, as well as novel multi-sensory device technologies. Smart wireless environments, enabled by the technology of Reconfigurable Intelligent Surfaces (RISs)~\cite{di2019smart,RIS_challenges,WavePropTCCN,risTUTORIAL2020,pan2022overview,zhi2022active,single_amplifier_george,RIS_Scattering,9827873}, constitute a recent revolutionary wireless connectivity paradigm contributing to the latter vision, according to which, the propagation of information-bearing ElectroMagnetic (EM) signals can be dynamically programmed over the air in a cost-, power-, and computationally-efficient manner~\cite{RISE6G_COMMAG}. 

Recent theoretical investigations on the joint optimization of massive Multiple-Input Multiple-Output (MIMO) transceivers with RISs have demonstrated the potential of RIS-enabled smart wireless environments for improving spectral and energy efficiencies~\cite{huang2019reconfigurable,KSI+22,KDA22,10694582,9673796,10670007}, localization accuracy~\cite{RIS_Localization,10230036,9726785,8313072,10124713,STAR_RIS_Loc}, computing~\cite{RIS_Computing_2023,HMIMO_Computing}, physical-layer security~\cite{PLS_Kostas,Counteracting,10289918}, as well as the integration of sensing and communications~\cite{RIS_ISAC_SPM}. More specifically, in~\cite{You_2021}, focusing on a single-RIS-assisted multi-user MIMO system in the uplink direction and considering slowly varying statistical Channel State Information (CSI) availability between the user terminals and the RIS, an algorithm for joint passive and active beamforming optimization was proposed. A multi-user MIMO system aided by multiple RISs, with the objective to maximize the system sum rate with respect to the RISs' phase configurations and the beamforming vectors at the Transmitters (TXs), was studied in \cite{ADD21}. In \cite{Yang_2021c}, the problem of energy efficiency optimization for a wireless communication system assisted by multiple distributed RISs was investigated. RIS-empowered Device-to-Device (D2D) communications underlaying a cellular network were considered in~\cite{Yang_2021d}, in which an RIS was missioned to enhance the desired signals and suppress interference between paired D2D and cellular links. Targeting the uplink achievable sum-rate maximization in a D2D cellular system with multiple distributed RISs deployed at the cell boundaries in~\cite{CLN+21}, a low-complexity decentralized optimization algorithm was presented. In~\cite{Yang_2021c}, a system with multiple distributed RISs was investigated for energy efficiency, optimizing the dynamic control of the on-off status of each RIS as well as the phase profiles of the on-state RISs. An RIS-aided multi-user MIMO system with statistical CSI availability, aiming at sum-rate maximization, was studied in~\cite{ZMS+23}, while~\cite{CHP+23} focused on optimizing the uplink of an RIS-assisted wideband multi-user system operating in the near-field regime. A multiple access scheme for non-orthogonal multi-user access in wireless communication systems assisted by multiple RISs was presented in~\cite{9693982}. Multi-RIS-enabled intelligent wireless channels were also studied in \cite{Samarakoon_2020,pervasive_DRL_RIS}, with their optimization being performed using, respectively, supervised and unsupervised artificial intelligence methodologies.

Another category of research works on RIS-enabled smart wireless environments deals with the characterization of their fundamental capacity limits. In particular,~\cite{Mu_2021b,LSC+23} focused on the downlink of an RIS-assisted multi-user wireless communication system, with the latter work considering millimeter-wave channels under the assumption of statistical CSI availability. In~\cite{PTR+22}, the authors analyzed the achievable sum rate of the RIS-aided MIMO broadcast channel showcasing the advantages of deploying multiple RISs, and presented optimization algorithms for the passive and active beamforming optimization. {More recently, the authors in \cite{chen2023fundamental} took advantage of the duality between uplink and downlink to characterize the dirty-paper-coding capacity region of a multi-antenna TX broadcasting to multiple single-antenna Receivers (RXs) in the presence of a single RIS, and optimized the system via zero-forcing transmission.} An asymptotic analysis of the uplink data rate in a multi-user setup with a single RIS acting as a receiver was carried out in~\cite{Jung2020}, considering spatially correlated Rician fading channels subject to estimation errors as well as RIS hardware impairments. For the same system model, but for the downlink direction,~\cite{Nadeem2020} studied the optimum linear precoding matrix that maximizes the minimum signal-to-interference-plus-noise ratio, and presented deterministic approximations for its parameters. An asymptotic analysis and approximations for the interference-to-noise ratio considering uncorrelated Rayleigh fading channels were derived in~\cite{INR_analysis}, while~\cite{analysis_intertwinement} presented upper and lower bounds for the outage probability and ergodic capacity, considering the intertwinement model of 
\cite{Abeywickrama_2020_all} for the amplitude and phase relation in the RIS element response. In 
\cite{Gao2021_irs_train_outage}, the authors generalized the performance metric to the outage probability, in the context of high speed trains. The achievable sum capacity of an RIS-aided multi-user MIMO system was also analyzed in~\cite{JS23}, and two efficient transmission schemes were designed. The effect of a single RIS on the uplink transmission from multiple TXs to a single multi-antenna RX was analyzed in~\cite{You_2021}, obtaining an asymptotic expression for the sum Mutual Information (MI) averaged over the channel matrices only close to the RIS, thus, neglecting the fast-fading effects close to the TXs and RX. To the best of our knowledge,~\cite{Moustakas2023_RIS} and its conference version were the first works analyzing the asymptotic outage capacity of a multi-RIS-enabled smart wireless environment assisting a point-to-point Kronecker-correlated MIMO system.

\subsection{Chapter's Contributions}
In this chapter, the capacity of MIMO-MAC-RIS communication systems, comprising multiple multi-antenna TXs simultaneously communicating with a single multi-antenna RX in the presence of multiple RISs, is analyzed. In particular, closed-form expressions for the asymptotic statistics of the sum-MI performance are presented, which are used to assess the potential gains from the RISs, but also their limitations. A summary of the contributions of this work appear below:
\begin{itemize}
\item Applying methods used originally in the context of statistical physics, a closed-form expression for the average sum-MI with arbitrary relative priorities for each multi-antenna TX in the presence of multiple RISs is presented. This expression, when optimized over the phase configuration matrices of the RISs, provides the capacity region for the MIMO-MAC-RIS problem. The expression is nominally valid in the limit of large numbers for the TXs/RX antenna elements and large numbers for the phase-tunable elements of the RISs.

\item Using the above methods, an analytic expression for the variance of the sum-MI is also presented. It is discussed why the higher cumulant moments of the distribution of the sum-MI vanish in the asymptotic limit, as a result of which the distribution of the sum-MI converges weakly to the Gaussian distribution. Thus, an approximation for its outage probability for block-fading channels is derived. This approximation is corroborated using Monte Carlo simulations, where the agreement with the approximate Gaussian distribution is exceptional down to outage probabilities of $10^{-3}$, even for relatively small antenna arrays at both the TXs and RX.

\item By considering the ergodic sum-MI as a metric of the MIMO-MAC-RIS system performance, the phase configurations of the multiple RISs are designed via two novel approaches, a semi-optimal and an optimal one, assuming that only the statistical properties of the involved channels are known, which is more realistic, given their slow variations compared to the fading of the channels. The optimization with respect to the RISs' reflections using the analytic expressions of the ergodic sum-MI performance is far more efficient compared to optimizing the instantaneous sum-MI. Furthermore, due to a decoupling effect, the optimization over the multiple RISs can be performed at separate steps, hence, it can be easily parallelized.

\item Using the derived ergodic sum-MI expression, the optimization gains over the phase configurations of the multiple RISs is analyzed and the throughput of the optimal solution is compared to the semi-optimal one. Furthermore, the effectiveness of RISs' phase optimization as the number of TXs increases is assessed, observing that beyond a certain point the relative gains vanish. 

\item Finally, the numerically evaluated analytic results are compared with Monte Carlo simulations, demonstrating very good agreement even when relatively small antenna arrays at the TXs and RX are considered. The effect of phase quantization of the RIS tunable elements to the overall performance of the MIMO-MAC-RIS system is also evaluated. It is showcased that, even with an $1$-bit phase quantization, the optimization gains are significant, while with a $2$-bit quantization, the performance is nearly optimal.
\end{itemize}

\subsection{Chapter's Outline}
Section~\ref{sec:MIMO channel model} describes the proposed system and channel models, while Section~\ref{sec:MI_Analysis} introduces the closed-form expressions for the first two cumulant moments of the sum-MI and the  MIMO-MAC-RIS capacity region. Section~\ref{sec:MI_Optimization} showcases the optimization methodology of the asymptotic sum-MI performance and capacity regions as a function of the phase matrices of the multiple RISs. In Section~\ref{sec:Numerical_Results}, considering a number of different system settings, the analytic results are compared with numerically generated ones after optimizing over the phase matrices. Finally, in Section~\ref{sec:Conclusions}, the concluding remarks of the chapter are included. 

\subsection{Chapter's Notations} 
Bold-faced upper-case letters denote matrices, e.g., $\vec X$, with its $(i,j)$-element expressed as $[\vec X]_{i,j}$, while bold-faced lower-case letters stand for column vectors, e.g., $\vec x$, with its $i$-element given by $[\vec x]_i$. The superscripts $T$ and $\dagger$ indicate
transposes and Hermitian conjugates, $\Tr{\, \cdot\,  }$ and $\left\Vert\cdot\right\Vert_F$ represent the trace and Frobenius norm of a matrix, respectively, whereas $\vec I_n$ ($n\geq2$) is the $n$-dimensional identity matrix. The superscripts/subscripts $t$ and $r$ denote quantities (e.g., channel matrices) corresponding to the TX and RX, respectively. Finally, $\mathbf{x}\sim{\cal CN}(\mathbf{0}_n,\mathbf{I}_n)$ represents an $n$-element complex and circularly symmetric Gaussian vector with zero-mean elements and covariance matrix $\mathbf{I}_n$, while $\ex[\,\cdot\,]$ is the expectation operator.

\begin{figure}[!t]
	\centering
	\includegraphics[width=\columnwidth]{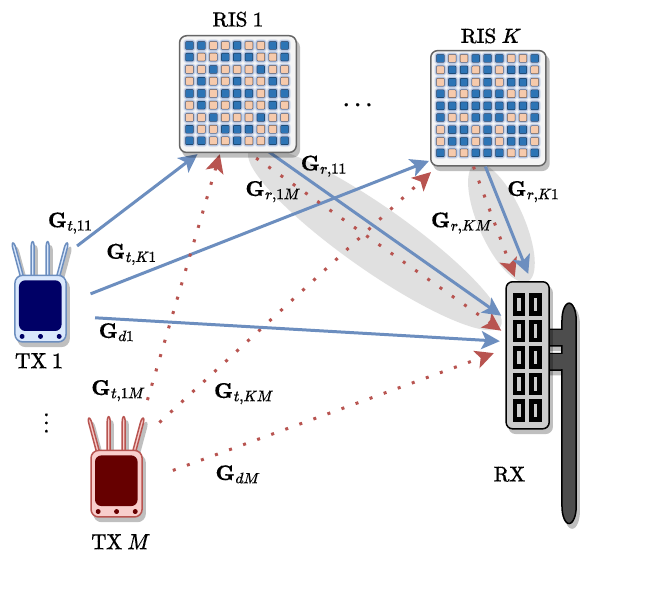} 
	  \caption{
   The considered MIMO-MAC-RIS communication system comprising $\nue$ TXs, each equipped with $\nt$ antenna elements, a single $\nr$-antenna RX, and $K$ identical RISs. The $\bG_{\alpha,\beta}$ notations in the figure represent the channel gain matrices between any pair $\alpha$ and $\beta$ of the latter network nodes.
   }
		\label{fig:system_model}
\end{figure}
\section{System and Channel Models}\label{sec:MIMO channel model}
\subsection{System Model}
The wireless communication system of Fig.~\ref{fig:system_model} comprises $\nue$ TXs, each equipped with $\nt$ antenna elements, which are simultaneously communicating with a single $\nr$-antenna RX via the assistance of $\nris$ identical RISs, each consisting of $\ns$ tunable reflecting elements~\cite{Tsinghua_RIS_Tutorial}. The fading channel under which this MIMO-MAC-RIS system operates includes the $\nue$ direct TX-RX links as well as channel components resulting from the programmable reflections due to the multiple RISs in the TX-RX vicinity. All channel gains are assumed to be perfectly available at the RX via an adequate estimation approach (e.g.,  \cite{Tsinghua_RIS_Tutorial,hardware2020icassp,Swindlehurst_CE,HRIS_CE_all,zhi2022power}), but not to any of the TXs. 

\subsection{Received Signal Model}
The $\nr$-dimensional received signal vector at the RX in baseband representation is given by the following mathematical expression:
\begin{equation}\label{eq:basic_channel_eq}
  \by \triangleq \sum_{m=1}^\nue \bG_{{\rm tot},m}  \bx_m + \bz,
\end{equation}
where $\bz\sim{\cal CN}(\mathbf{0}_\nr,\mathbf{I}_\nr)$ represents the thermal noise vector and $\bx_m$, with $m=1,2,\ldots,M$, is the $\nt$-dimensional signal vector transmitted from the $m$-th TX having signal covariance matrix $\bQ_m\triangleq \ex[\bx_m\bx_m^\dagger]$, which has the following normalization $\Tr{\bQ_m}=\rho\nt$. The parameter $\rho$ denotes the common Signal-to-Noise Ratio (SNR) of all TXs.
Each $\nr\times\nt$ matrix of channel amplitudes $\bG_{{\rm tot},m}$  in \eqref{eq:basic_channel_eq} between RX and each $m$-th TX can be mathematically expressed as follows:
\begin{align}\label{eq:Gtot}
\bG_{{\rm tot},m} \triangleq \bG_{dm} + \sum_{k=1}^\nris\bG_{r,k}\bPhi_k\bG_{t,km},
\end{align}
with $\bG_{r,k}$ and $\bG_{t,km}$, for $k=1,2,\ldots,\nris$, denoting the $\nr\times \ns$ and $\ns\times\nt$ channel matrices between the RX and each $k$-th RIS and between that $k$-th RIS and each $m$-th TX, respectively. $\bG_{dm}$ is  the $\nr\times\nt$ matrix with elements the direct channel gains between the RX and each $m$th TX. Note that this channel does not include any reflection from any of the RISs. In addition, each $\bPhi_k$ denotes the $\ns$-dimensional diagonal square matrix containing the tunable reflection coefficients at each $k$-th RIS in the main diagonal. In particular, each $\bPhi_k$'s non-zero elements are modeled to have unit norm \cite{risTUTORIAL2020}, with each $n$-th reflection coefficient, with $n=1,2,\ldots,\ns$, of each $k$-th RIS expressed as $[\bPhi_k]_{n,n}\triangleq e^{i\phi_{k,n}}$. 

\subsection{Channel Model}
The characteristics of the channel matrices are modeled using the so-called Kronecker-product correlation model~\cite{4786505,4786505,6184250}. Specifically, all channel matrices in $\bG_{{\rm tot},m}$ are assumed complex Gaussian, independent for different values of $m$, and have the following Kronecker-product-type covariances $\forall i,j=1,2,\ldots,\nr$, $\forall \ell,n=1,2,\ldots,\nt$, and $\forall a,b=1,2,\ldots,\ns$:
\begin{subequations}\label{eq:all_correlations}
\begin{align}
\label{eq:Gd_cov}
\ex\left[[\bG_{dm}]_{i,\ell}[\bG_{dm}]_{j,n}^*\right]&=\frac{\snrdir}{\nt}[\bR_{dm}]_{i,j} [\bT_{dm}]_{\ell,n},
\\ \label{eq:Grk_cov}
\ex\left[[\bG_{r,k}]_{i,a}[\bG_{r,k}]_{j,b}^*\right]&=\frac{1}{\nt}[\bR_{k}]_{i,j} [\bS_{r,{k}}]_{a,b},
\\ \label{eq:Gtk_cov}
\ex\left[[\bG_{t,km}]_{a,\ell}[\bG_{t,km}]_{b,n}^*\right]&=\frac{1}{\nt}[\bS_{t,km}]_{a,b} [\bT_{km}]_{\ell,n}.
\end{align}
\end{subequations}
The signal-impinging-side correlation matrices in the latter expressions, namely $\bR_{k}$, $\bS_{r,k}$, and $\bR_{dm}$, as well as the outgoing-signal-side correlation matrices, namely $\bT_{km}$, $\bS_{t,km}$, and $\bT_{dm}$ are all non-negative definite, with traces  fixed to: $\Tr{\bT_{km}}=\Tr{\bT_{dm}}=\nt$, $\Tr{\bR_{k}}=\Tr{\bR_{dm}}=\nr$, and $\Tr{\bS_{r,k}}=\Tr{\bS_{t,km}}=\ns$. Note that the elements of the matrix $\bS_{t,km}$ represent the correlation coefficient between the incoming EM waves from the $m$-th TX at the 
elements of the $k$-th RIS,
while the elements of $\bS_{r,k}$ denote the correlation between outgoing (reflected) EM waves from the $k$-th RIS to the RX. Finally, $\snrdir$ represents the ratio of the SNR of the direct link between each $m$-th TX and RX over the SNR of the link between the same nodes realized via the $k$-th RIS. 

Similar to \cite{Moustakas2000_BLAST1_new,Moustakas2023_RIS}, each $(a,b)$-element of each correlation matrix $\bS_{t,km}$ in \eqref{eq:all_correlations} can be obtained as follows:
\begin{equation}\label{eq:corr_mat_w(k)_def}
    \left[\bS_{t,km}\right]_{ab}=\int \, w_{t,km}(\bk) \exp\left(i\bk^T\left(\bx_a-\bx_b\right)\right)d\Omega_{\bk},
\end{equation}
where $w_{t,km}(\bk)$ represents a normalized weight function of the power of the incoming EM wave from direction $\bk$, with the latter denoting the corresponding three-dimensional wave vector with magnitude $|\bk|=k_0\triangleq\frac{2\pi}{\lambda}$, where $\lambda$ is the wavelength. Note that this function can be characterized by the mean direction of arrival $\bs_0$ (with $|\bs_0|=k_0$) and the Angle Spread (AS) $\sigma$ (in radians) via the following expression~\cite{Moustakas2023_RIS}:
\begin{equation*}\label{eq:weight_fn_def}
    w_{t,km}(\bk)\propto \exp\left(-\frac{|\bk-\bs_0|^2}{2\sigma^2k_0^2}\right).
\end{equation*}
The three-dimensional vectors $\bx_a$ and $\bx_b$ in \eqref{eq:corr_mat_w(k)_def} denote the positions of the meta-material elements $a$ and $b$ within the $k$-th RIS. The integral in this expression is evaluated over the whole unit sphere, and it is normalized so that, for $\bx_a=\bx_b$, it gives unity. A similar expression can be given for all correlation matrices in \eqref{eq:all_correlations}.

\section{Capacity Analysis}
\label{sec:MI_Analysis}
In the following, the usual assumption that the RX knows the overall end-to-end channel matrices $\bG_{{\rm tot},m}$ $\forall$$m$ given in \eqref{eq:Gtot} (through, e.g., pilot-assisted channel estimation~\cite{HRIS_Mag_all,Tsinghua_RIS_Tutorial, Swindlehurst_CE}) is made. Since all channels fluctuate due to fast fading conditions, the system's long-term performance is captured through the ergodic averages of the sum-MI over the fading distribution. For convenience, the ergodic sum-MI performance for the considered MIMO-MAC-RIS system is defined as follows:
\begin{align}
    \label{eq:I(Q,Phi)}
    &I\left(\{\bQ_m\}_{m\in{\cal S}},\{\bPhi_k\}_{k=1}^K\right) 
    \triangleq \ex\left[\log\det\left(    
    \vec I_\nr + \sum_{m\in {\cal S}}\bG_{{\rm tot},m} \bQ_m \bG_{{\rm tot},m}^\dagger
    \right)\right],
\end{align}
where the averaging is over the channel matrices $\bG_{{\rm tot},m}$'s for ${\cal S} \subseteq \{1,2,\ldots,M\}$. In this expression, the dependences on the input covariance matrices $\bQ_m$'s as well as the phase configurations matrices $\bPhi_k$'s for all RISs are explicitly shown. The above expression represents the ergodic sum-MI for the set ${\cal S}$ of TXs being active, while all others being silent.

\subsection{MIMO-MAC without RISs}
In the absence of RISs (hence, setting $\bPhi_k=\mathbf{0}_{\ns}$ $\forall$$k$ in~\eqref{eq:I(Q,Phi)}), the capacity region of a pure MIMO-MAC system has been well analyzed in the past \cite{Cheng1993_GaussianMAC_ISI_CapacityRegion, Tse1998_GaussianMAC1_PolymatroidStructure, Vishwanath2001_OptimumMAC, Goldsmith2003_CapacityLimitsMIMO, Yu2004_IterativeMIMOMAC},
and consists of all rate multiplets $\{R_1,R_2,\ldots,R_M\}$ such that it holds:
\begin{align}
\label{eq:cap_region_def}
    \sum\limits_{m\in {\cal S}} 
    R_m \leq I\left(\{\bQ_m\}_{m\in{\cal S}}\right)
\end{align}
for all sets ${\cal S}\subseteq \{1,2,\ldots,M\}$. In this expression, for simplicity, the sum-MI's dependence on $\bPhi_k$'s has been excluded, since they are all equal to the identity matrix. It is well known that the above capacity region is achieved with Gaussian inputs at each TX, where each covariance matrix $\bQ_m$ at the each $m$-th TX is constrained such that $\Tr{\bQ_m}\leq \nt$ holds.

The covariance matrices that correspond to the points on the boundary of the capacity region can be obtained by maximizing the following functional with respect to all $\bQ_m$'s~\cite{Yu2004_IterativeMIMOMAC}:
\begin{align}\label{eq:functional_pure}
    &{\cal L}\left(\{\bQ_m\}_{m=1}^M,\boldmu\right)\triangleq
    \mu_M I\left(\{\bQ_m\}_{m=1}^M\right) 
+\sum_{\ell=1}^{M-1}\left(\mu_{\ell}-\mu_{\ell+1}\right) 
    I\left(\{\bQ_m\}_{m=1}^\ell\right) 
\end{align}
for all possible non-negative vectors $\boldmu\triangleq[\mu_1,\mu_2,\ldots,\mu_M]$ of relative priorities given to each of the TXs, and for a fixed sequential interference cancellation order, where it was assumed without loss of generality that $\mu_1\geq \mu_2\geq \ldots \mu_M$. The optimization is performed given the statistical knowledge of the channel matrices (covariances matrices) appearing in \eqref{eq:all_correlations}. 

\subsection{MIMO-MAC Assisted by RISs}
Let the channel matrices $\bG_{{\rm tot},m}$'s, for $m=1,2,\ldots,\nue$, be composed as in \eqref{eq:Gtot}, including the diagonal matrices $\bPhi_k$'s, for $k=1,2,\ldots,K$, with the tunable reflection coefficients for each of the $K$ RISs, and the input signal covariance matrices $\bQ_m$'s for each of the $M$ TX antenna arrays. By extending the previous result for the conventional MIMO-MAC case, the MIMO-MAC-RIS capacity region consists of all rate multiplets $\{R_1,R_2,\ldots,R_M\}$ such that it holds using \eqref{eq:I(Q,Phi)} for all sets ${\cal S} \subseteq \{1,2,\ldots M\}$ with $\Tr{\bQ_m}\leq \nt$ $\forall$$m$:
\begin{align}
\label{eq:cap_region_prop}
    \sum\limits_{m\in {\cal S}} 
    R_m \leq I\left(\{\bQ_m\}_{m\in{\cal S}},\{\bPhi_k\}_{k=1}^K\right).
\end{align}
The matrices $\bQ_m$'s and $\bPhi_k$'s corresponding to the borders of this region are obtained by maximizing the following functional: 
\begin{align}\label{eq:Lagrangian_mu_cap}
    L_M\!\left(\{\bQ_m\}_{m=1}^M,\{\bPhi_k\}_{k=1}^K,\boldmu\right)\triangleq&  
    \mu_M I\left(\{\bQ_m\}_{m=1}^M,\{\bPhi_k\}_{k=1}^K\right)\nonumber
\\ 
&+\sum_{\ell=1}^{M-1}\left(\mu_{\ell}-\mu_{\ell+1}\right) 
    I\left(\{\bQ_m\}_{m=1}^\ell,\{\bPhi_k\}_{k=1}^K\right)
\end{align}
for all possible non-negative vectors $\boldmu$ as in \eqref{eq:functional_pure} of relative priorities given to each TX, and for a fixed sequential interference cancellation order, where it has been again assumed, without loss of generality, that $\mu_1\geq \mu_2\geq \ldots\geq \mu_M$ and $\sum_{\ell=1}^M\mu_\ell=1$. 

The quantity $I(\{\bQ_m\}_{m=1}^M,\{\bPhi_k\}_{k=1}^K)$ appearing in~\eqref{eq:Lagrangian_mu_cap} needs to be evaluated for fixed input covariance matrices $\bQ_m$'s and RIS reflection matrices $\bPhi_k$'s. The authors in~\cite{MA_TWC_2024}, assuming that the matrices $\bG_{dm}$ and $\bG_{t,km}$'s for the direct channel to RX and the channel matrices from each of the $k$-th RIS, respectively, have zero-mean complex Gaussian elements with covariances given by \eqref{eq:Grk_cov}, \eqref{eq:Gd_cov}, and \eqref{eq:Gtk_cov}, respectively, showed that, when $\nt, \nr, \ns\to\infty$ with fixed ratios $\beta_r\triangleq\nr/\nt$ and $\beta_s\triangleq\ns/\nt$, $I(\{\bQ_m\}_{m=1}^M,\{\bPhi_k\}_{k=1}^K)$ normalized per TX antenna element can be expressed as follows: 
\begin{align}\label{eq:S0}
& C_M\left(\{\bQ_m\}_{m=1}^M,\{\bPhi_k\}_{k=1}^K\right) \triangleq \frac{\ex[I\left(\{\bQ_m\}_{m=1}^M,\{\bPhi_k\}_{k=1}^K\right)]}{\nt}
\nonumber  \\ %
=&\frac{1}{\nt}\log\det \left(\bI_\nr  + \tilde{\bR} \right)
+\frac{1}{\nt} \sum_{k=1}^K\sum_{m=1}^M \log\det \left(\bI_{\ns} +  
t_{1k}r_{2km}\bSigma_{km}\right)
 \\%
  &+\frac{1}{\nt}\sum_{m=1}^\nue\log\det \left(\bI_\nt + 
  \bQ_m \tilde{\bT}_m  \right)
- \sum_{m=1}^M\left(r_{dm} t_{dm}+\sum_{k=1}^K\left(r_{1km}t_{1k}+r_{2km}t_{2km}\right)\right).\nonumber
\end{align}
In the above expression, the matrices $\tilde{\bR}$, $\tilde{\bT}_m$, and $\bSigma_{km}$ are defined as: 
\begin{align}
    \tilde{\bR}&\triangleq 
    \sum_{m=1}^\nue\left( r_{dm} \bR_{dm} + \sum_{k=1}^\nris r_{1km}\bR_{k}\right),
    \label{eq:R_tilde}\\
    \tilde{\bT}_m&\triangleq 
    t_{dm}\bT_{dm}+  \sum_{k=1}^\nris t_{2km}\bT_{km},
    \label{eq:T_tilde}\\
    \bSigma_{km} &\triangleq 
    \bS_{t,km}^{1/2}\bPhi_k^\dagger\bS_{r,k}\bPhi_k\bS_{t,km}^{1/2}
    \label{eq:Sigma_k_initial},
\end{align}
where the quantities $t_{dm}$, $t_{2km}$, $r_{dm}$, $r_{2km}$, $r_{1km}$,  and $t_{1k}$ can be evaluated by solving the fixed point equations as below:
\begin{align}
\nonumber
    t_{dm} &=\frac{1}{\nt}\Tr{\left(\bI_\nr + \tilde{\bR} \right)^{-1}\bR_{dm}},\,\,
    t_{2km} = \frac{t_{1k}}{\nt} \Tr{\left(\bI_{\ns} +  
t_{1k}r_{2km}\bSigma_{km}  \right)^{-1} \bSigma_{km}}\\
    \nonumber
    r_{dm} &= \frac{1}{\nt} \Tr{\left(\bI_\nt + \bQ_m
    \tilde{\bT}_m
    \right)^{-1}\bQ_m\bT_{dm}}, \,\,
    r_{2km} = \frac{1}{\nt} \Tr{\left(\bI_\nt + 
    \bQ_m \tilde{\bT}_m
    \right)^{-1}\bQ_m\bT_{km}},\\ \nonumber
     r_{1km}& = \frac{r_{2km}}{\nt} \Tr{\left(\bI_{\ns} +  
t_{1k}r_{2km}\bSigma_{km}\right)^{-1}  \bSigma_{km}},\,\,t_{1k} =\frac{1}{\nt}\Tr{\left(\bI_\nr   + \tilde{\bR}  \right)^{-1}\bR_{k}}.
\end{align}    
In addition, the variance of $I(\{\bQ_m\}_{m=1}^M,\{\bPhi_k\}_{k=1}^K)$ takes the limiting form:
\begin{align}\label{eq:Var(I)}
 {\rm Var}(I(\{\bQ_m\}_{m=1}^M,\{\bPhi_k\}_{k=1}^K))\triangleq-\log\det({\bf \Lambda}),
\end{align}
where the $(2M+4MK)$-dimensional matrix $\bLambda$ is given by: 
\begin{align}\label{eq:Vmat_def}
    {\bf \Lambda} = \left[\begin{array}{cccccc} 
    {\bf M}_{dt} & 0 & {\bf M}_{2dt}^T & 
    -\bI_{\nue} & 0 & 0
    \\   0 & {\bf M}_{1t} & 0 & 
    0 & -\bI_{\nue\nris} & {\bf M}_{12}
    \\
    {\bf M}_{2dt} & 0 & {\bf M}_{2t} & 
    0 & 0 & -\bI_{\nue\nris}
    \\
    -\bI_\nue & 0 & 0 & 
    {\bf M}_{dr} & {\bf M}_{1dr}^T & 0
    \\
    0 & -\bI_{\nue\nris} & 0 & 
    {\bf M}_{1dr} & {\bf M}_{1r} & 0
    \\
    0 & {\bf M}_{12} & -\bI_{\nue\nris} & 
    0 & 0 & {\bf M}_{2r}
    \end{array}\right].
\end{align}
In the latter expression, the matrices ${\bf M}_{1r}$, etc., are defined as follows:
\begin{align}
    &[{\bf M}_{1r}]_{k,m}^{k'm'} \triangleq -\frac{1}{\nt} \Tr{ \overline{{\bf R}} {}^{-1}  {\bf R}_{km}\overline{{\bf R}} {}^{-1} {\bf R}_{k'm'}}, 
 \\ 
    &[{\bf M}_{2r}]_{k,m}^{k'm'} \triangleq -\delta_{kk'}\delta_{mm'}\frac{t_{1k}^2}{\nt} \Tr{ \left(\overline{{\bf S}}_{km}\right)^{-2}  {\bf \Sigma}_{km}^2 },
    \\ 
    &[{\bf M}_{1t}]_{k,m}^{k'm'} \triangleq -\delta_{kk'}\delta_{mm'}\frac{r_{2km}^2}{\nt} \Tr{ \left(\overline{{\bf S}}_{km}\right)^{-2}  {\bf \Sigma}_{km}^2 },
    \\ 
    &[{\bf M}_{2t}]_{k,m}^{k'm'} \triangleq  -\frac{\delta_{mm'}}{\nt}\Tr{ \left(\overline{{\bf T}}_m\right)^{-1}  {\bf Q}_m{\bf T}_{mk}\left(\overline{{\bf T}}_m\right)^{-1} {\bf Q}_m{\bf T}_{mk'}},\\
    &[{\bf M}_{12}]_{k,m}^{k'm'} \triangleq \delta_{kk'}\delta_{mm'}\frac{1}{\nt} \Tr{ \left(\overline{{\bf S}}_{km}\right)^{-2}  {\bf \Sigma}_{km} },\\
    &[{\bf M}_{1dr}]_{k,m}^{m'} \triangleq -\frac{1}{\nt} \Tr{ \overline{{\bf R}} {}^{-1}  {\bf R}_{km}\overline{{\bf R}} {}^{-1} {\bf R}_{dm'}},\\
    &[{\bf M}_{2dt}]_{k,m}^{m'}\triangleq -\frac{\delta_{mm'}}{\nt}\Tr{ \left(\overline{{\bf T}}_m\right)^{-1}  {\bf Q}_m{\bf T}_{km}\left(\overline{{\bf T}}_m\right)^{-1} {\bf Q}_m{\bf T}_{dm}},
    \\ 
    &[{\bf M}_{dt}]_{m,m'} \triangleq -\frac{1}{\nt} \Tr{ \left(\left(\overline{{\bf T}}_m\right)^{-1}  {\bf Q}_m{\bf T}_{dm}\right)^2},\\
    &[{\bf M}_{dr}]_{m,m'}\triangleq  -\frac{1}{\nt} \Tr{ \overline{{\bf R}} {}^{-1}  {\bf R}_{dm} \overline{{\bf R}} {}^{-1}  {\bf R}_{dm'}},
    \end{align}
where the matrix notations $\overline{{\bf R}} \triangleq {\bf I}_{\nr}+\tilde{\bf R}$, $\overline{{\bf T}}_m\triangleq{\bf I}_{\nt}+{\bf Q}_m\tilde{\bf T}_m$, and $\overline{{\bf S}}_{km}\triangleq{\bf I}_{\ns}+t_{1k}r_{2km}\bSigma_{km}$ have been used.

By inspecting~\eqref{eq:S0}, it can be seen that the matrices $\bPhi_k$'s, appearing through the quantities $\bSigma_{km}$, depend only on the incoming and outgoing covariance matrices $\bS_{r,k}$'s and $\bS_{t,km}$'s, and not on other covariance matrices of the problem. Furthermore, as also noted in \cite{Moustakas2023_RIS}, the $\bPhi_k$'s of the different RISs are decoupled, since they appear in separate $\log\det(\cdot)$ terms. This allows for separate optimization of the phases of each matrix $\bPhi_k$. The underlying reason for this feature lies on the fact that, as it can be seen in \eqref{eq:basic_channel_eq}, the RIS phase configuration matrices appear ``sandwitched'' between independent channel matrices. Since the phases of these matrices are independent, it is easy to see that they can be optimized independently. In the asymptotic limit, the effect of all other phase configuration matrices enters into the optimization of a given $\bPhi_k$ through the scalar quantities $r_{1km}$, $t_{1k}$, $r_{2km}$, $t_{2km}$, $r_{dm}$, and $t_{dm}$, which only depends on all other phase matrices and in an aggregate manner. In addition, the $\bSigma_{km}$'s, each corresponding to a different $m$-th TX, also appear in separate $\log\det(\cdot)$ terms. As it will be seen below, this will enable faster optimization of the $\bPhi_k$'s. It is finally noted that, combining \eqref{eq:cap_region_prop} and~\eqref{eq:S0}, yields the ergodic capacity region per antenna in the asymptotic limit of large numbers of TXs/RX antennas and RIS reflecting elements.

It is finally noted that the variance of $I(\{\bQ_m\}_{m=1}^M,\{\bPhi_k\}_{k=1}^K)$ provides a metric for its variability. For example, it can be observed that, while this metric is of order $O(\nt)$, its variance is of order unity. Furthermore, since all higher moments can be shown to vanish in the large $\nt$ limit, the distribution of $I(\{\bQ_m\}_{m=1}^M,\{\bPhi_k\}_{k=1}^K)$ can be shown to be asymptotically Gaussian. Note that, using the theoretical framework developed in~\cite{Tse1998_GaussianMAC1_PolymatroidStructure}, one can also obtain the outage capacity region for the above problem. This outage capacity region can be calculated more easily in the asymptotic limit, by noting that the joint probability distribution of $I(\{\bQ_m\}_{m=1}^M,\{\bPhi_k\}_{k=1}^K)$, for different sets ${\cal S}$ can be shown, using tools from statistical physics \cite{Moustakas2003_MIMO1}, to be jointly Gaussian. 

\section{MIMO-MAC-RIS Capacity Optimization}\label{sec:MI_Optimization}
Capitalizing on the asymptotic ergodic sum-MI performance analysis of MIMO-MAC-RIS systems in the previous section, the optimization of the ergodic sum-MI with respect to the phase configuration of the multiple RISs, namely the elements of all $\bPhi_k$'s, is now discussed. The purpose of the optimization is twofold. First, it results to the maximum ergodic sum-capacity of such systems, which is a metric indicating the maximum total throughput for the case of $\nue$ TXs and $\nris$ RISs. In addition, it provides the boundaries of the ergodic MIMO-MAC capacity region in the presence of multiple RISs. As mentioned in the previous section, the focus will be only in the case where there is no need to optimize the signal covariance matrices $\bQ_m$'s, for example, when the covariance matrices at all $M$ TXs, i.e., $\bT_{km}$'s and $\bT_{dm}$'s, are identity matrices. Hence, following the system model in~\eqref{eq:Gtot} and the geometry depicted in Fig.~\ref{fig:system_model}, and using the capacity region in~\eqref{eq:Lagrangian_mu_cap} as well as the closed-form asymptotic sum-MI expression in~\eqref{eq:S0}, the following Optimization Problem (OP) formulation is considered:
\begin{align*} 
\begin{split}
    \mathcal{OP}_1: \,\, \max_{\{\bPhi_k\}_{k=1}^\nris} &
    \,\,\Bigg[\mu_M C_M\left(\{\bPhi_k\}_{k=1}^K\right)+\left.\sum_{\ell=1}^{M-1}\left(\mu_{\ell}-\mu_{\ell+1}\right) 
    C_\ell\left(\{\bPhi_k\}_{k=1}^K\right)\right]
    \\ 
    & \hspace{-0.4cm}\text{s.t.} \quad \lvert [\bPhi_k]_{n,n} \rvert = 1  \, \, \forall k,n.
\end{split}
\end{align*}
As previously discussed, the ergodic sum-MI in~\eqref{eq:S0} is simpler, when compared to the corresponding exact expression in~\eqref{eq:I(Q,Phi)}, since the 
RISs' phase configuration matrices $\bPhi_k$'s appear in separate logarithms. This property simplifies the solution of $\mathcal{OP}_1$; interestingly, the asymptotic sum-MI optimization can be performed separately for each $\bPhi_k$, i.e., for each $k$-th RIS.

To simplify the exposition of the methodology, only the case of the sum capacity, namely, when $\mu_M=1$ and $\mu_1=\ldots=\mu_{M-1}=0$, will be presented. In this case, to solve the resulting $\mathcal{OP}_1$ form, an alternating optimization approach~\cite{J:alternating_minimization} is deployed, according to which, at each algorithmic iteration, the values of the variables $r_{1km}$, $t_{1k}$, $r_{2km}$, $t_{2km}$, $r_{dm}$, and $t_{dm}$ are initially fixed and the sum capacity is maximized over the $\bPhi_k$'s. As previously mentioned, it is possible to separate the process into  independent ones over different $\bPhi_k$'s. Hence, the multi-RIS phase configuration design problem can be made simpler as follows. For each $k$-th RIS, making use of~\eqref{eq:S0}, the following OP needs to be solved:
\begin{align*} 
\begin{split}
    \mathcal{OP}_2: \max_{\bPhi_k}\sum_{m=1}^M &\log\det \left(\bI_{\ns} + t_{1k}r_{2km}\bPhi_k^\dagger\bS_{r,k}\bPhi_k\bS_{t,km} \right)
\\    \hspace{0.4cm}
\text{s.t.} \quad &\lvert [\bPhi_k]_{n,n} \rvert = 1  \, \, \forall k,n.
\end{split}
\end{align*}
Once $\mathcal{OP}_2$ is solved for each of the $\nris$ RISs, all obtained phase matrices to calculate the parameters $r_{1km}$, $t_{1k}$, $r_{2km}$, $t_{2km}$, $r_{dm}$, and $t_{dm}$ are substituted. These two steps may be repeated iteratively, until all parameters converge to their optimal values, or until the number of iterations exceeds its maximum value.

Two related ways to obtain the optimum $\bPhi_k$'s, each solving one of the $\nris$ separate $\mathcal{OP}_2$ problems, with both being based on the gradient ascent approach, will be henceforth presented. The first, in Section~\ref{sec:Analytical_Solution} that follows, is a semi-optimal algorithm, which, while, not converging at a maximum, is less computationally intensive and performs well at low signal ASs, as will be demonstrated numerically in the following section with the performance evaluation results. Subsequently, in Section \ref{sec:Numerical_Solution}, the full gradient ascent approach solving $\mathcal{OP}_2$ will be described.

\subsection{Semi-Optimal Optimization over $\bPhi_k$'s}
\label{sec:Analytical_Solution}  
To better understand the form of the sum-capacity-optimal solution for $\bPhi_k$'s, the focus will be first on a MIMO-MAC-RIS system with one RIS, dropping for convenience the RIS numbering index $k$. In addition, the limit of vanishing AS, for which the matrices $\bS_{t,m}$'s and $\bS_{r}$ are effectively of unit rank, corresponding to a single incoming plane wave to the RIS from each $m$-th TX and a single outgoing plane wave from the RIS, respectively, will be analyzed. In~\cite{Moustakas2023_RIS}, it was shown that the RIS phase configuration solutions for this vanishing AS approximation case can very accurately agree with the full optimization analysis for general AS values. This will be similarly confirmed, in the following numerical results' section, for the MIMO-RIS-MAC focus in this chapter. Hence, let us commence by expressing $\bS_{t,m}=\ns \bu_m\bu_m^\dagger$ $\forall$$m$ and $\bS_{r}=\ns \bv\bv^\dagger$,
where the unit-norm vectors $\bu_m$ and $\bv$ correspond to the wave vectors of the incoming signal from the $m$-th TX and the outgoing signal to the RX and have components, each having the elements:
\begin{align}
    \label{eq:S_eigenvector}
    \left[\bu_m\right]_n = \frac{1}{\sqrt{\ns}} e^{i\bq_{t,m}\bx_n}, \quad
    \left[\bv\right]_n = \frac{1}{\sqrt{\ns}} e^{i\bq_{r}\bx_n},
\end{align}
where $\bx_n$ represents the location of the $n$-th RIS element, and $\bq_{t,m}$ and $\bq_{r}$ are the incoming wave-vectors from the $m$-th Tx to the RIS and the outgoing wave vector from the RIS to the RX, respectively. For this case where the AS is vanishing, the matrices $\bSigma_m$'s can be expressed as follows:
\begin{align}
\label{eq:unit_rank_Sigma_m}
    \bSigma_m= N_s^2 |\kappa_{m}(\bPhi)|^2 \bu_{m}\bu^\dagger_{m},
\end{align}
where the scalar parameter $\kappa_{m}$ is given by:
\begin{align}
\label{eq:kappa_m}
\kappa_{m}(\bPhi)&\triangleq\bv^\dagger\bPhi\bu_{m}=\frac{1}{\ns}\sum_{n=1}^{\ns} e^{i\left(\phi_n-\Delta\bq_m\bx_n\right)}
\end{align}
with $\Delta\bq_m\triangleq\bq_{r}-\bq_{tm}$. By substituting~\eqref{eq:S_eigenvector} into~\eqref{eq:unit_rank_Sigma_m} and then the resulting expression into the previously derived asymptotic ergodic sum-MI formula, the second term in~\eqref{eq:S0} can be written as follows (for the case of a single RIS):
\begin{align}\label{eq:C_M}
  D_M(\bPhi) \triangleq  \sum_{m=1}^M \log\left(1 + N_s^2t_{1m}r_{2m}|\kappa_m(\bPhi)|^2\right).
\end{align}
It is noted that, in the case of only one TX in the system, i.e., when $M=1$, the summation in~\eqref{eq:C_M} diminishes to one term including $\kappa_1(\bPhi)$, and it can be optimized over this parameter when the modulus of $\kappa_1(\bPhi)$ is maximal, i.e., when $\phi_n=\Delta\bq\bx_n$ $\forall$$n$; this has been recently shown in~\cite{Moustakas2023_RIS}. However, in the general MIMO-MAC-RIS case with $M>1$ TXs, each of the phase components $\phi_n$ needs to be chosen to optimize all $\kappa_m(\bPhi)$'s simultaneously, as shown from the $M$ terms included in~\eqref{eq:C_M}. It is clear that, the more separated the quantities $\Delta\bq_m$'s for different values of $m$ are, the higher the competition of the terms in \eqref{eq:C_M}'s summation will be, and hence, the smaller the values of $D_M(\bPhi)$ in~\eqref{eq:S0} will be. 

To optimize over the RIS phase configuration $\bPhi$ (i.e., solve $\mathcal{OP}_2$ for the vanishing AS case), gradient ascent on the above functional is performed. It can be shown that, at each $i$-th algorithmic iteration, the reflection coefficient at each $n$-th RIS element in the diagonal of $\bPhi^{(i)}$ is updated as follows: 
\begin{align}\label{eq:GD_iteration_simple}
    &\phi_n^{(i)}=\phi_{n}^{(i-1)}+\epsilon\sum_{m=1}^M{\rm Im}\left(\frac{\ns t_{1m}r_{2m}\kappa_m\left(\bPhi^{(i-1)}\right) e^{i\left(\Delta \bq_m \bx_n-\phi_n^{(i-1)}\right)} }{1+N_s^2t_{1m}r_{2m}\left|\kappa_m\left(\bPhi^{(i-1)}\right)\right|^2}  \right),
\end{align}
where $\epsilon$ is a parameter chosen to make the algorithm converge fast. When the above iteration converges, e.g., at the $\mathcal{I}$-th iteration, the obtained matrix $\bPhi^{(\mathcal{I})}$ can be used to compute the fixed point values of the parameters $r_{1km}$, $t_{1k}$, $r_{2km}$, $t_{2km}$, $r_{dm}$, and $t_{dm}$, which are then fed back into~\eqref{eq:C_M} until the whole algorithm converges. 

Note that the same procedure can be employed to obtain the borders of the capacity region using~\eqref{eq:cap_region_prop}, \eqref{eq:Lagrangian_mu_cap}, and~\eqref{eq:S0}. In particular, in this case, for fixed $\boldmu$, gradient ascent can be adopted to maximize the following functional with respect to the RIS phase configuration matrix $\bPhi$: 
\begin{align}\label{eq:cap_mu12}
  D_M(\bPhi) \triangleq  \sum_{\ell=1}^M (\mu_\ell-\mu_{\ell+1})\!\sum_{m=1}^\ell \log\left(1 + N_s^2t_{1m}^\ell r_{2m}^\ell |\kappa_m(\bPhi)|^2\right)\!,
\end{align}
where, for simplicity, $\mu_{M+1}=0$ was set in~\eqref{eq:Lagrangian_mu_cap}. In the above, the superscripts on the parameters $t_{1m}^\ell$ and $r_{2m}^\ell$ are included to specify that these correspond for each $\ell$-value to the fixed parameters $r_{1km}$, $t_{1k}$, $r_{2km}$, $t_{2km}$, $r_{dm}$, and $t_{dm}$ for the case of $\ell$ TXs.

\subsection{Optimal Solution for $\bPhi_k$'s}\label{sec:Numerical_Solution} 
In the general case of arbitrary values for the angular spread, the full gradient of the functional objective appearing in $\mathcal{OP}_2$ with respect to each of the RIS phase configuration matrices needs to be evaluated. Similar to the treatment in Section~\ref{sec:Analytical_Solution}, the focus is made on a single $\bPhi$. Clearly, solving $\mathcal{OP}_2$ is computationally more complex, due to the necessity to invert a large $\ns\times\ns$ matrix. Specifically, at each $i$-th algorithmic iteration, the reflection coefficient at each $n$-th RIS element in the diagonal of $\bPhi^{(i)}$ needs to be updated as follows: 
\begin{align}\label{eq:GD_iteration_full}
    &\phi_n^{(i)}=\phi_{n}^{(i-1)}
    +\epsilon\sum_{m=1}^M{\rm Im}
    \left[
    \left(\bI_{\ns} + t_{1m}r_{2m}\left(\bPhi^{(i-1)}\right)^\dagger\bS_{r}\bPhi^{(i-1)}\bS_{t,m} \right)^{-1}
    \right]_{n,n},
\end{align}
where the values of the parameters $t_{1m}$ and $r_{2m}$ are fed from the previous iteration. In principle, and as mentioned above, one should perform the above iteration until it converges, for example, at the $\mathcal{I}$-th iteration, and then use the resulting $\bPhi^{(\mathcal{I})}$ matrix (for every RIS) to re-evaluate the parameters $t_{1m}$ and $r_{2m}$. However, since the equations for $r_{1km}$, $t_{1k}$, $r_{2km}$, $t_{2km}$, $r_{dm}$, and $t_{dm}$ also involve the evaluation of the inverse of the matrix appearing above, it is more computationally efficient to jointly perform the optimization of the parameters and the RIS phase configuration matrix. The procedure for obtaining the optimal solution of the sum-MI metric in $\mathcal{OP}_1$, i.e., when $\mu_M=1$ and $\mu_1=\ldots=\mu_{M-1}=0$, for multiple RISs is summarized in Algorithm~\ref{alg:loc} (top of the next page). This approach has been found to be numerically stable, especially when compared to the algorithm of~\cite{Zhang_Capacity}, and similar ones, that perform algorithmic iterations per RIS element. Also, the same convergence values were tested and found with several initial conditions (for example, all zeros or random initial conditions). The overall algorithmic steps for solving the same problem via the semi-optimal approach of Section~\ref{sec:Analytical_Solution} can be obtained in a similar manner.
\begin{algorithm}[!t]
    \caption{Sum-MI Optimizing $\bPhi_k$'s for the MIMO-MAC-RIS}
    \label{alg:loc}
    \begin{algorithmic}[1]
        \renewcommand{\algorithmicrequire}{\textbf{Input:}}
        \renewcommand{\algorithmicensure}{\textbf{Output:}}
        \REQUIRE $\epsilon$, $\delta$, $\mathcal{I}$, $\bR_{dm}$, $\bT_{dm}$, $\bR_{k}$, $\bS_{r,{k}}$, $\bS_{t,km}$, and $\bT_{km}$ $\forall$$k,m$. 
        \STATE Initialize $r_{1km}^{(0)}=t_{1k}^{(0)}=r_{2km}^{(0)}=t_{2km}^{(0)}=r_{dm}^{(0)}=t_{dm}^{(0)}=0$ and $\bPhi_k^{(0)}=\bI_\ns$ $\forall$$k,m$.
        \FOR{$i=1,2,\ldots,\mathcal{I}$ }
        \FOR{$k=1,2,\ldots,\nris$ }
                \STATE Set $\mathbf{A}_{km}^{(i)}\!=\!t_{1km}^{(i-1)}r_{2km}^{(i-1)}\bS_{t,km}\left(\!\bPhi_k^{(i-1)}\!\right)^\dagger\bS_{r,k}\bPhi_k^{(i-1)}$ $\forall$$m$. 
                \STATE Compute $\mathbf{B}_{km}^{(i)}\!=\!\left(\bI_{\ns}+\mathbf{A}_{km}^{(i)}\right)^{-1}$
                .\STATE Use matrix $\mathbf{B}_{km}^{(i)}$ to update $r_{1km}^{(i)}$, $t_{2km}^{(i)}$, $t_{dm}^{(i)}$, $t_{1km}^{(i)}$, $r_{dm}^{(i)}$, and $r_{2km}^{(i)}$.
                \FOR{$n=1,2,\ldots,\ns$ }
                \STATE Set $\phi_{k,n}^{(i)}=\phi_{k,n}^{(i-1)}+\!\epsilon\!\sum_{m=1}^M{\rm Im}\left[\mathbf{B}_{km}^{(i)}\right]_{n,n}$.
                \ENDFOR
        \ENDFOR
                \IF{$\left\lVert\bPhi_{k}^{(i)}-\bPhi_{k}^{(i-1)}\right\rVert+  
                \left|r_{1km}^{(i)}-r_{1km}^{(i-1)}\right|+\left|t_{2km}^{(i)}-t_{2km}^{(i-1)}\right|+\left|t_{dm}^{(i)}-t_{dm}^{(i)}\right|+\left|t_{1km}^{(i-1)}-t_{1km}^{(i-1)}\right|+\left|r_{dm}^{(i)}-r_{dm}^{(i-1)}\right|+\left|r_{2km}^{(i)}-r_{2km}^{(i-1)}\right|
                <\delta$ }
                    \STATE Output the optimized $\bPhi_k^{(i)}$ and break.
        \ENDIF
        \ENDFOR
        \STATE Output each optimized $\bPhi_k^{(\mathcal{I})}$.
    \end{algorithmic}
\end{algorithm}

\section{Numerical Results and Discussion}\label{sec:Numerical_Results}
In this section, numerically evaluated results on the statistics of the sum-MI performance for the considered MIMO-MAC-RIS system between multiple TX arrays and a single RX array are presented. To focus on the effects of the presence of the RISs, the direct paths between the TXs and the RX have been neglected. In addition, the fading conditions in the vicinity of the transmit and receive antennas were assumed to be uncorrelated. 

\begin{table}[!t]
\caption{Simulation Parameters used in Figs.~\ref{fig:MI_AS_2UE}--\ref{fig:cap_region}.}
\label{table1}
\centering
\begin{tabular}{|c|c||c|c|}
\hline
\textbf{Parameter} &\textbf{Value} & \textbf{Parameter} &\textbf{Value}\\
\hline \hline
Carrier frequency & $2.5$GHz & Wavelength $\lambda$& $12$ cm \\
\cline{1-2} \cline{3-4}
Direct TX-RX SNR & $0$ dB & SNR $\rho$ & $10$ dB \\
\cline{1-2} \cline{3-4}
RIS elements $\ns$ & $400$ & RIS inter-element distance & $6$ cm  \\
\hline
Elevation $\theta_{\rm in}$ & $30^{\circ}$ &Elevation $\theta_{out}$ & $70^{\circ}$ \\
\hline
TX Antennas $\nt$ & 4 & RX Antennas $\nr$ & 8 \\
\hline
\end{tabular}
\end{table}
\begin{figure}[!t]
	\centering
	\includegraphics[width=1.05\columnwidth]{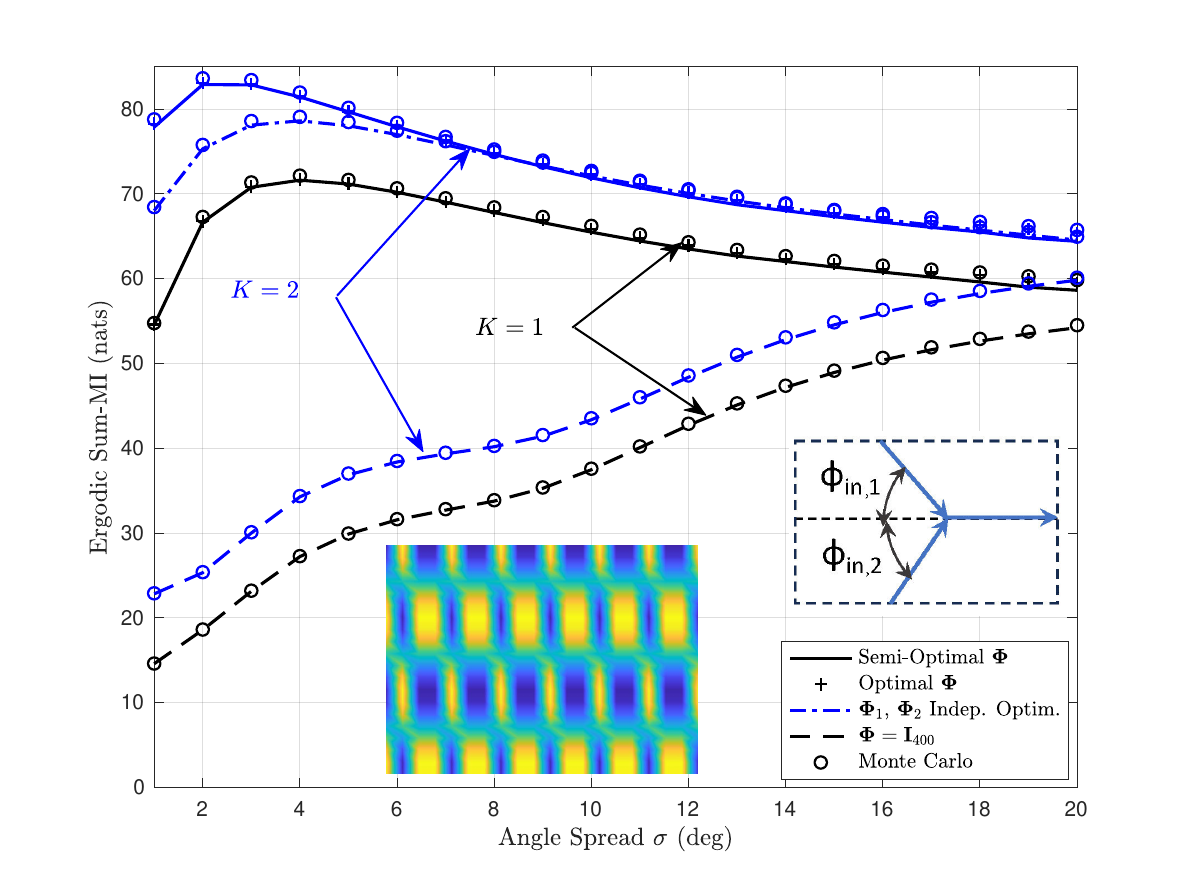}
	  \caption{The ergodic sum-MI performance in nats per channel use for $M=2$ TXs and for $K=1$ and $K=2$ RISs versus the angle spread $\sigma$ in degrees. All other parameter values of this simulation setup appear in Table~\ref{table1}, except for the incoming signal azimuth angles from each TX to the RISs, which are set as $\phi_{{\rm in},1}=45^{\circ}$ and $\phi_{{\rm in},2}=-45^{\circ}$, as can be seen in the inset. The lower left inset figure depicts the optimal phase distribution on the RIS when the angle spread is equal to $\sigma=5^{\circ}$.}
		\label{fig:MI_AS_2UE}
\end{figure}
Figure~\ref{fig:MI_AS_2UE} depicts the ergodic sum-MI performance versus the AS $\sigma$ with $K=1$ and $K=2$ RISs, each having $400$ elements located in a square grid of $\lambda/2$ inter-element spacing, in the presence of $M=2$ TX arrays. The two TX arrays have different incoming azimuth angles at the RISs, set to $\phi_{in,1}=45^o$ and $\phi_{in,2}=-45^o$, respectively, as seen in the inset. All other parameter values of this simulation setup appear in Table~\ref{table1}. In this figure, the black curves correspond to single RIS performance ($K=1$), while the blue curves correspond to $K=2$ RIS performance. Furthermore, the dashed curves were obtained with the unoptimized phase configuration ${\bf \Phi}={\bf I}_{400}$, while the solid curves correspond to semi-optimal ${\bf \Phi}$, obtained through the analysis of Section \ref{sec:Analytical_Solution}. The crosses are obtained applying the fully-optimal methodology of Section~\ref{sec:Numerical_Results}. Finally, all circles correspond to the expectation of the MI evaluated through Monte Carlo simulations. It can be observed that the semi-optimal and fully optimal solutions perform identically, and that all analytic solutions are quite accurate, falling on top of the Monte Carlo simulations. 
It is clear that, even in the case of a single RIS, the gain due to RIS optimization is substantially larger for small AS values. This is not surprising since smaller ASs correspond to higher correlations, and thus optimization based on the covariance matrix of the channel is better performing. Of course, the performance gains for $K=2$ RISs are also larger for smaller AS values. 

However, given that the optimization for each phase matrix ${\bf \Phi}_k$ for each $k$-th RIS necessitates the knowledge of the matrices ${\bf S}_{r,k}$ and ${\bf S}_{t,km}$ for all TXs $m$ (here $m=1,2$), it is interesting to entertain the idea of optimizing ${\bf \Phi}_k$ from the information of only one TX, essentially pairing each RIS with a given TX. In this case, as has been seen in \cite{Moustakas2023_RIS}, the optimization procedure can be much faster, and has been shown to be essentially equivalent to setting the phase of each element in ${\bf \Phi}_k$ with the difference of the phases of the corresponding elements of the eigenvectors with the largest eigenvalue of 
the matrices ${\bf S}_{r,k}$ and ${\bf S}_{t,km}$ (for $m=k$), which, from 
\eqref{eq:kappa_m} can be seen that it means:  
\begin{align}
\bPhi_{nn} = e^{i\phi_n} = e^{(\bq_r-\bq_{t,km})\bx_n}.
\end{align}
In this expression, $\bq_r$ and $\bq_{t,km}$ are the outgoing wave vectors from the $k$-th RIS and the incoming wavevector to the RIS with index $k=m$. In this case, the resulting throughput is the blue dash-dot curve with the ``Indep. Optim.'' term. As shown, the behavior is not as good as the optimally optimized one (solid blue), especially for small AS values. This is not surprising, since the phases of each RIS are optimized without taking into account the potential interference caused to the other TX. Nevertheless, the total throughput is higher than the one corresponding to a single RIS (black solid line). Finally, it is interesting to observe the optimal phase distribution on one of the RISs in the presence of both TXs on the lower left inset figure.

\begin{figure}[!t]
	\centering
	\includegraphics[width=1.05\columnwidth]{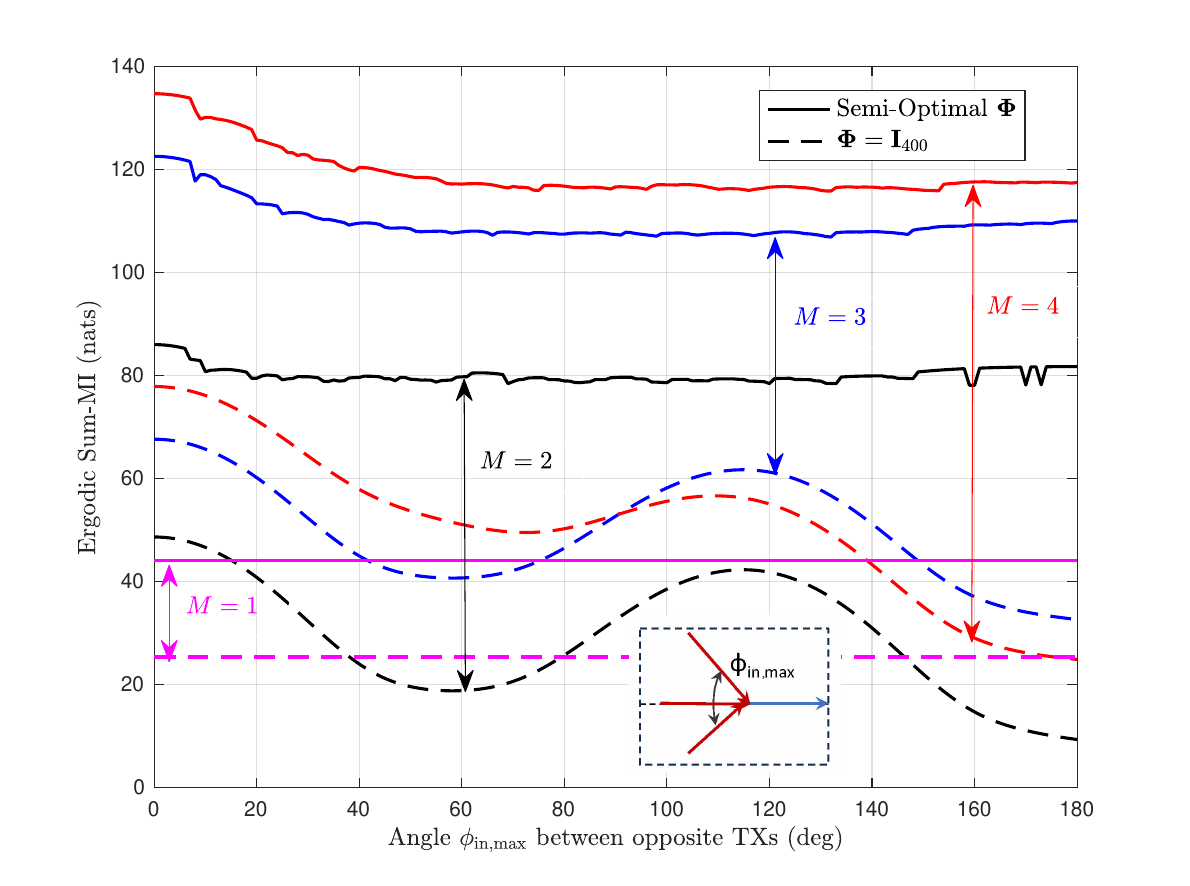} 
\caption{The ergodic sum-MI performance in nats per channel use in the presence of $K=1$ RIS for $\nue=2$ (black), $\nue=3$ (blue), and $\nue=4$ (red) TXs. The RX array has $\nr=12$ antennas, while each TX has $\nt=4$ antennas. All incoming and outgoing signals were considered to have angle spread $\sigma=4^{\circ}$. The remaining simulation parameters take values from Table~\ref{table1}. For the cases where $\nue>1$, the TXs have incoming azimuth angles that are equidistant with maximum angle equal to the one plotted on the $x$-axis of the plot. For concreteness, we have included an inset which shows the azimuth angles for the case of $\nue=3$. The solid lines correspond to the semi-optimal approach described in Section~\ref{sec:Analytical_Solution}. }
\label{fig:MI_phi_1234UE}
\end{figure}
In Fig.~\ref{fig:MI_phi_1234UE}, the following key question of the chapter is addressed: How many users can be served efficiently using a single RIS? In particular, the figure includes the ergodic sum-MI as a function of the total angular separation $\phi_{\rm in,max}$ between the $\nue$ incoming rays, separated by equal azimuth angles from each other, for $\nue=\{1,2,3,4\}$. It can be observed that the proposed semi-optimal optimization procedure for $\bPhi$ converges more robustly than the optimal optimization method, which is not plotted here. The reason is that, especially for large total angular separation, full gradient descent does not converge due to presence of multiple local minima of the algorithm. In contrast, the semi-optimal method focuses to the angular directions of the largest eigenvalues of the correlation matrices, and hence faces less difficulties to converge to the semi-optimum point. For the number of antennas of the RX chosen ($\nr=12$) versus the number of TX antennas per user, it can be seen the suggestive result that, up to $\nue=3$, the gains of the sum-MI are substantial, while diminishing for larger number of users ($\nue=4$). It is also shown that, for small $\phi_{in,max}$, the gains are larger, due to the fact that the optimal directivity properties for all TXs are the same and there is no competition between trying to satisfy different TXs.

\begin{figure}[!t]
\centering
\includegraphics[width=1.05\columnwidth]{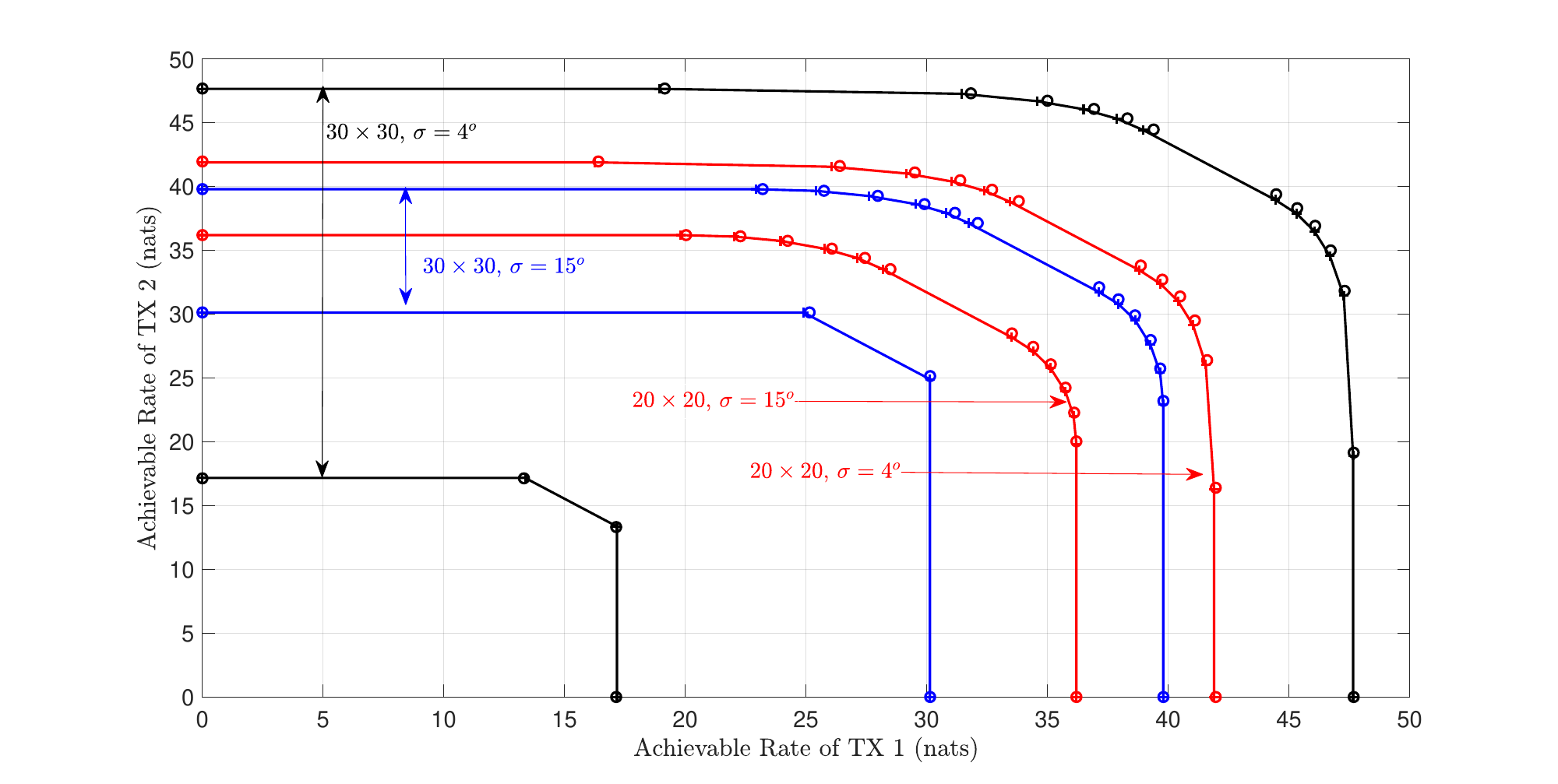}
\caption{Ergodic capacity region for the case of $\nue=2$ TXs and $K=1$ RIS, with incoming signal azimuth angles from each TX set to $\phi_{{\rm in},1}=45^{\circ}$ and $\phi_{{\rm in},2}=-45^{\circ}$, considering the angle spread values $\sigma=4^{\circ}$ and $\sigma=15^{\circ}$. The blue and black curves correspond to an RIS with $\ns=400$ elements, while the magenta and red ones correspond to $\ns=900$ elements, all arranged in a square grid of $\lambda/2$ inter-element spacing. All other parameter values are included in Table~\ref{table1}. 
}
\label{fig:cap_region}
\end{figure}
The effects of the optimization of a single RIS serving $M=2$ users on their capacity region for the two representative AS values $\sigma=4^o$ and $\sigma=15^o$ and two different sizes of square RISs ($\ns=20\times 20=400$ and $\ns=30\times 30=900$) are illustrated in Fig.~\ref{fig:cap_region}. It can be seen that, in the absence of any optimization (i.e., $\bPhi=\bI$) (lower two curves), the capacity boundary takes the usual form of a pentagon, with the low AS case having overall less capacity. In contrast, when the phase matrix $\bPhi$ is optimized, the capacity boundary is further away from the origin for smaller AS values, owing to higher beamforming gains. Clearly, larger RISs provide larger capacities for all ASs. Finally, for the case of optimal $\bPhi$, the capacity region is no longer a pentagon, but rather curved. The circles on each curve correspond to the rate pairs evaluated from Monte Carlo simulations, with the circles corresponding to the $\bPhi_1$ natrix obtained by numerically optimizing expression~\eqref{eq:Lagrangian_mu_cap} for $\mu_1=0:0.1:1$ and $\mu_2=1-\mu_1$.

\begin{figure}[!t]
\centering
\includegraphics[width=1.05\columnwidth]{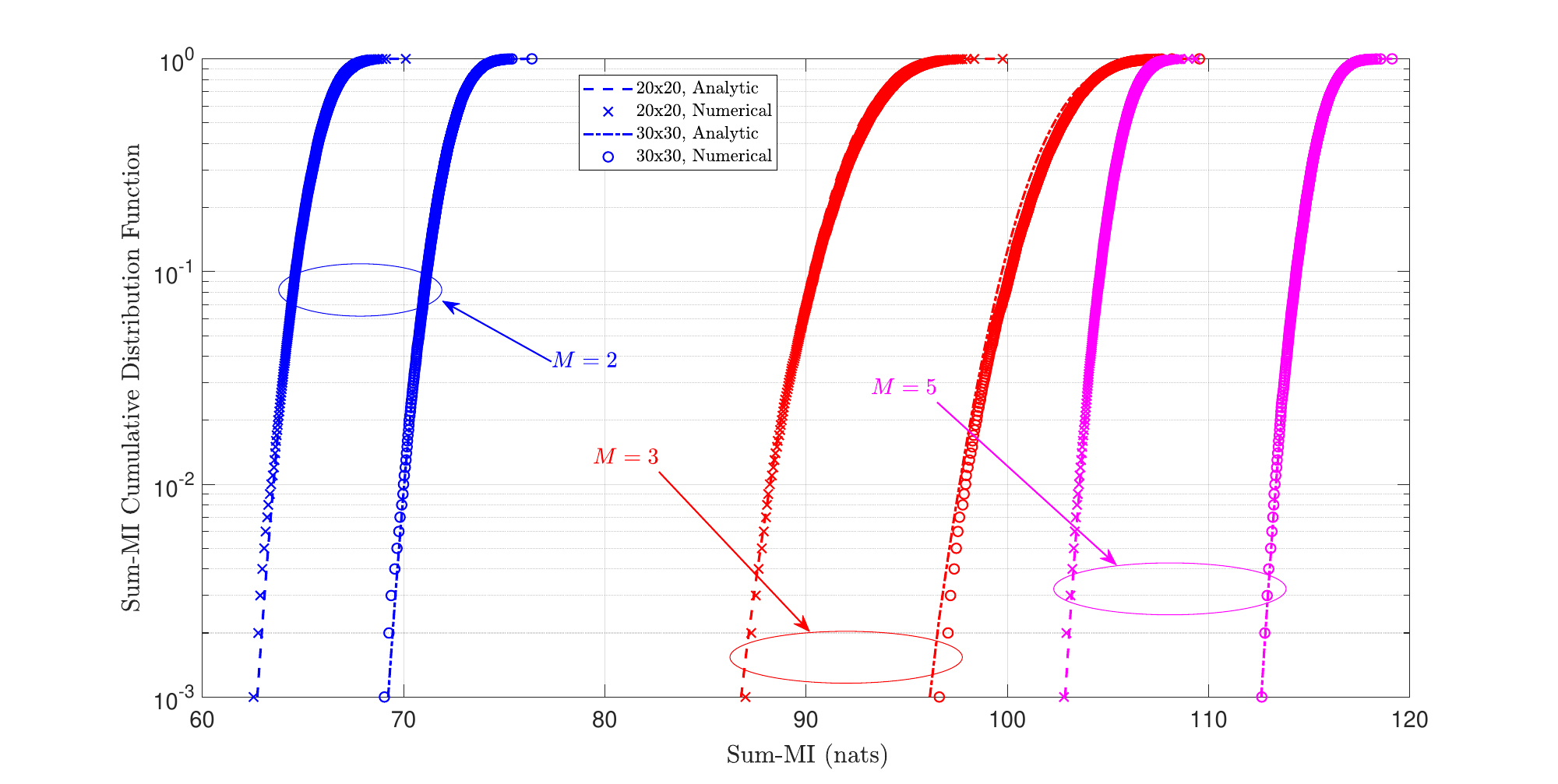}
\caption{The cumulative distribution of the sum-MI performance in nats with $\nris=2$ RISs and with $\nue=2$ (blue), $\nue=3$ (red), and $\nue=5$ (magenta) TXs, respectively. For each $\nue$ value, two different sizes of RISs were considered, namely $\ns=400$ (dashed lines and 'x' marks) and $\ns=900$ (dash-dot lines and 'o' marks) elements arranged in a square grid of $\lambda/2$ inter-element spacing. Additionally, a setting with $\nt=4$ and $\nr=12$ antennas and SNR of $\rho=10$ dB was selected. The solid curves (dash and dash-dot) correspond to a Gaussian approximation using the mean and variance obtained analytically in the chapter, and are compared with curves generated via Monte Carlo simulations ('x' and 'o' marks). }
\label{fig:MI_CDFs}
\end{figure}
Finally, the cumulative distribution of the sum-MI performance with $\nris=2$ RISs and various numbers of TXs, namely, $\nue=\{2,3,5\}$, and for two sizes of RISs, namely, $\ns=400$ and $\ns=900$, is plotted in Fig.~\ref{fig:MI_CDFs}. As observed, the solid curves, corresponding to a Gaussian approximation using the mean and variance obtained analytically in \eqref{eq:S0} and \eqref{eq:Var(I)}, agree remarkably well with curves generated via Monte Carlo simulations, all the way down to $10^{-3}$ outage probability. For simplicity, all channels were assumed to be uncorrelated and were set as $\bPhi_k=\bI_{\ns}$ $\forall$$k$. It can be seen that, for increasing $\nue$ and $\ns$, the gain in the median throughput as well as the variance (related to the average slope of the curves) is progressively decreasing. 

\section{Conclusion}\label{sec:Conclusions} 
In this chapter, methods from statistical physics were deployed to showcase the gains from the use of multiple RISs in the presence of multiple multi-antenna users transmitting individual signals to a single multi-antenna RX (i.e., the MIMO-MAC-RIS system). Analytic expressions for the statistics of the sum-MI performance for this MAC system were presented showcasing the benefits emanating from RISs. While the designed methodologies are in principle valid  when the numbers of antennas are large, it was demonstrated numerically that the presented analytic results are valid even for moderate antenna sizes, thus providing a method to analytically provide dimensioning for a multi-user network in the presence of multiple RISs. It was also shown that the optimization of the RIS phase configurations can be broken into separate optimizations, one for each RIS, using only statistical data of the nearby incoming and outgoing channels (i.e., their covariance matrices). This is especially true for highly correlated channels close to an RIS, for example due to a small AS value, which is reasonable for wireless systems with increasing carrier frequencies.

In addition, a semi-analytic optimization approach for the RISs which focuses only on the eigenvector of the correlation matrices with the largest eigenvalue was presented, which speeds up the RIS phase configuration design while also helping it to avoid non-relevant local maxima (minima). These approaches were compared with the case where each RIS is optimized to serve a single TX antenna, and it was found that, while a fully optimized RIS provides superior performance, a distributed local approach still provides significant benefits. This is something that needs to be taken into account when looking for complexity tradeoffs.

\bibliographystyle{IEEEtran}
\bibliography{references, wireless}
\end{document}